\documentclass[pmlr]{jmlr}

\usepackage{longtable}
\usepackage{ifthen}

 \usepackage{booktabs}
 \usepackage[flushleft]{threeparttable}
 \usepackage{color}
 \usepackage{soul}

\usepackage{lipsum}

\usepackage{enumitem}
\usepackage{comment}

\usepackage[load-configurations=version-1]{siunitx}

\makeatletter
\def\set@curr@file#1{\def\@curr@file{#1}}   
\makeatother

\theorembodyfont{\upshape}
\theoremheaderfont{\scshape}
\theorempostheader{:}
\theoremsep{\newline}

\jmlrvolume{219}
\jmlryear{2023}
\jmlrworkshop{Machine Learning for Healthcare}

\newcommand{\dataset}{EWAL\xspace}
\newcommand{\Dataset}{\textbf{E}D \textbf{W}riting \textbf{A}udit \textbf{L}ogs\xspace}

\title[Conceptualizing ML for Dynamic Information Retrieval of EHR Notes]{Conceptualizing Machine Learning for Dynamic Information Retrieval of Electronic Health Record Notes}

\author{\Name{Sharon Jiang}\hspace{0.5mm}$^{1}$
       \Email{jiangs@mit.edu}\\
       \Name{Shannon Shen}\hspace{0.5mm}$^{1}$
       \Email{zjshen@mit.edu}\\ 
       \Name{Monica Agrawal}\hspace{0.5mm}$^{1}$
       \Email{monica.n.agrawal@gmail.com}\\ 
       \Name{Barbara Lam}\hspace{0.5mm}$^{2,3}$
       \Email{blam@bidmc.harvard.edu}\\ 
       \Name{Nicholas Kurtzman}\hspace{0.5mm}$^{4}$
       \Email{nkurtzma@bidmc.harvard.edu}\\ 
       \Name{Steven Horng}\hspace{0.5mm}$^{3,4}$
       \Email{shorng@bidmc.harvard.edu}\\ 
       \Name{David Karger}\hspace{0.5mm}$^{1}$
       \Email{karger@mit.edu}\\ 
       \Name{David Sontag}\hspace{0.5mm}$^{1}$
       \Email{dsontag@csail.mit.edu}
       \\ \\
\addr{\textsuperscript{1}{\textit{Department of Electrical Engineering \& Computer Science, MIT, Cambridge, MA, USA}}}\\
\textsuperscript{2}{\textit{Division of Hematology \& Oncology, Department of Medicine, Beth Israel Deaconess Medical Center, Boston, MA, USA}}\\
\textsuperscript{3}{\textit{Division of Clinical Informatics, Department of Medicine, Beth Israel Deaconess Medical Center, Boston, MA, USA}}\\
\textsuperscript{4}{\textit{Department of Emergency Medicine, Beth Israel Deaconess Medical Center, Boston, MA, USA}}}

\begin{document}

\maketitle

\begin{abstract}

The large amount of time clinicians spend sifting through patient notes and documenting in electronic health records (EHRs) is a leading cause of clinician burnout.
By proactively and dynamically retrieving relevant notes during the documentation process, we can reduce the effort required to find relevant patient history.
In this work, we conceptualize the use of EHR audit logs for machine learning as a source of supervision of note relevance in a specific clinical context, at a particular point in time.
Our evaluation focuses on the dynamic retrieval in the emergency department, a high acuity setting with unique patterns of information retrieval and note writing. 
We show that our methods can achieve an AUC of 0.963 for predicting which notes will be read in an individual note writing session. 
We additionally conduct a user study with several clinicians and find that our framework can help clinicians retrieve relevant information more efficiently. Demonstrating that our framework and methods can perform well in this demanding setting is a promising proof of concept that they will translate to other clinical settings and data modalities (e.g., labs, medications, imaging).

\end{abstract}

\section{Introduction}

 Electronic health records (EHRs) serve as a central repository of a patient's past medical history, containing both structured data and free text notes ~\citep{burton2004using, LI2022100511}. These data are crucial across multiple stages of the medical decision making process \citep{Muhiyaddin2022}. Over the course of a patient encounter, clinicians describe time-varying purposes of information retrieval from the EHR: rapid sense-making of a new patient, re-familiarizing oneself with an existing patient, searching for a particular factoid, and looking for unspecified evidence to support the differential diagnosis process  \citep{nygren1992reading}. 
 
 However, retrieval of this relevant information is a time-consuming process, given the volume of data in EHRs. This data gathering process is complicated by the fact that much of the information required for medical decision making is found only in free text notes \citep{li2008comparing, shivade2014review}. These notes have become bloated and unwieldy to manually sift through due to their multitude of aims (clinical communication, compliance, and billing). Consequently, clinicians are spending more time navigating and documenting in EHRs than in face-to-face encounters with patients, a phenomenon considered a leading cause of clinician burnout \citep{menachemi2011benefits, moy2021measurement}. Given the inefficiencies in clinical workflows, there is significant interest in better characterizing and improving them.

In our work, we first present an in-depth analysis of the EHR note-writing \emph{process} based on 48,192 note-taking sessions in the emergency department. Prior work primarily focuses on analyzing high-level action sequences (i.e., record review, orders, documentation)~\citep{Rule2020-tu}, and few have looked into the granular activities of reading and writing during the documentation process. We then propose a framework that updates retrieval suggestions dynamically as the clinical context (e.g., the note being written) changes, thereby addressing the complex and time-varying information needs of clinical providers. Furthermore, we experiment with machine learning methods to support this proactive information retrieval of unstructured clinical notes to facilitate the writing process.  Much of the existing work in clinical natural language processing has focused on the analysis of static completed notes \citep{shivade2014review}. In our experiments, we show that our framework can achieve an AUC of 0.963 for a binary classification task to predict the most relevant notes during individual sessions.
Finally, we conduct a user study with clinicians and verify that our framework can help clinicians retrieve relevant information more efficiently.

\subsection*{Generalizable Insights about Machine Learning in the Context of Healthcare}

In this work, we focus on dynamically retrieving \textit{unstructured data} as a function of the current clinical context. However, this framework of dynamic information retrieval is also relevant across other modalities in the electronic health record, (e.g., labs, medications, imaging). This suggests a more general paradigm of learning proactive interfaces from EHR log data. Furthermore, due to large language models, we are at an inflection point in clinical natural language processing, at which computer-assisted clinical note generation is becoming more realizable. Consequently, it is important to take a step back to understand the workflows, information retrieval, and clinical reasoning that actually lead to the final documentation, rather than treating the note as a monolithic entity.

\section{Related Work}

\paragraph{ML for EHR Note Extraction and Retrieval} 
Many machine learning methods have been developed to help digest the critical information from unstructured free-text notes~\citep{LI2022100511}, e.g., extracting structured data~\citep{wang2018clinical} or summarizing key information~\citep{macavaney2019ontology, alsentzer2018extractive} from existing notes. 
To assist clinicians with navigating the high volumes of available EHR documentation, \citet{wei2018embedding} have explored using document embeddings~\citep{alsentzer-etal-2019-publicly} to retrieve relevant diagnostic ICD codes for a given document. 
Recently, large language models have demonstrated strong capabilities for understanding clinical text and can be used for a variety of tasks to process clinical notes \citep{agrawal2022large}.
However, most of this work focuses on a static setting, i.e., processing a fixed document. Our work explores how to utilize machine learning to dynamically retrieve relevant notes at different points during the creation of a note.

\paragraph{Characterization of EHR Activity with Audit Logs} 
One useful byproduct of EHRs is the data contained in \textit{audit logs}, which capture granular information about clinicians' activity (e.g., clicks, page views) within the EHR~\citep{Adler-Milstein2020-wz}. 
While initially intended for access control, audit logs have also been recently leveraged (i) to characterize the frequency and duration of different activities within the EHR and (ii) to mine higher-level patterns of activity sequences and clusters~\citep{Rule2020-tu}. 
Such information can inform a redesign of the user interfaces of EHRs to make existing workflows less clunky~\citep{Zheng2009-jy}. 
As audit logs can miss important context, such work has been supplemented by time-motion studies~\citep{Mamykina2012-nb, Rule2020-tu}. Unlike this previous work, we analyze not just the patterns of information and retrieval and documentation, but also the \textit{text content} of the notes and the design of predictive algorithms.

\paragraph{Proactive Information Retrieval} Given the high information needs of clinicians, there have been multiple efforts to develop support for dynamic information needs during clinical documentation.  
For example, Harvest~\citep{hirsch2015harvest} and Doccurate~\citep{ sultanumDoccurateCurationBasedApproach2019} use natural language processing (NLP) to help summarize, visualize, and navigate the patient's medical history, but they do not take into account the current clinical context in deciding what to display. 
\citet{King2019-ja} and \citet{gopinath2020fast} use learning-based methods to selectively highlight important structured data but not other free-text notes.
Our work differs from these in that we focus on the task of retrieving relevant \emph{unstructured EHR notes} during the documentation process.
A more recent system, MedKnowts~\citep{murray2021medknowts}, can surface other text notes if they contain relevant terms from a list created algorithmically. We focus on exploring with ML to better support proactive document retrieval.

\begin{figure}[t]
    \centering
    \includegraphics[width=\linewidth]{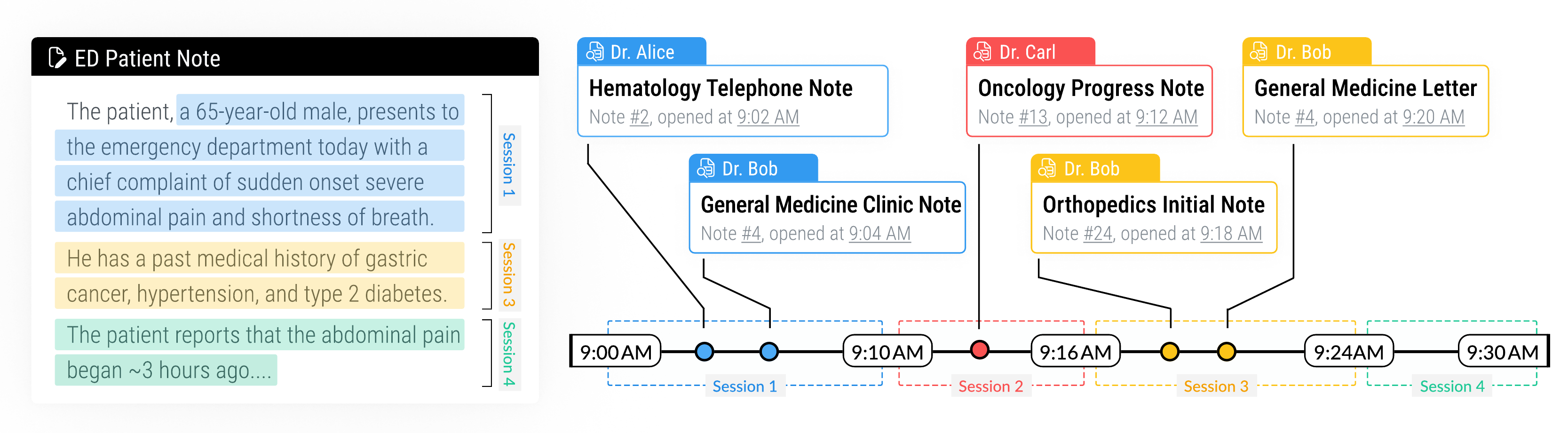}
    \caption{{\small Illustration of the documentation process after a patient presents to ED. We show the written note in the ED Patient Note panel (left) as well as the corresponding timeline for the information retrieval and writing (right). When a patient presents to the ED, a team of doctors starts to search for relevant past notes about the patient. The opening of each note is marked with dots on the timeline. The doctors then add written text in the ED Patient Note panel, and the EHR system saves the writing periodically; the time of each written snapshot is specified by the rounded rectangles on the timeline. We define the time between written snapshots as \textit{sessions}, and we color the added written text in the ED Patient Note panel based on the session when the writing was added. The same color is used to highlight the retrieved notes for a session. A total of four sessions occurred during the patient’s ED visit from 9 AM to 9:30 AM. There are sessions involving both retrieving past notes and writing the current note (e.g., session 1 and 3), or either of them (e.g., session 2 for retrieval or session 4 for writing exclusively). The ED Patient Note is based on text generated by GPT-4, and the writing time and IDs of the retrieved notes are synthetically generated.} }
    \label{fig:writing-timeline}
    \vspace{-5mm}
\end{figure}

\section{Understanding the ED Documentation Process with Audit Logs}

\subsection{The \Dataset (\dataset) Dataset}

While note taking happens at different points in the patient care process, we focus on the documentation process that occurs after a patient presents to the emergency department (ED). 
ED clinicians often see 35 patients in one shift, during which they must rapidly synthesize an often previously unknown patient's medical records in order to reach a tailored diagnosis and treatment plan \citep{murray2021medknowts}.
They are also in constant communication with their colleagues in other parts of the hospital in order to get specialty input and determine the best disposition for a patient. 
Because of the high acuity, high throughput nature of this setting, it may surface \emph{unique patterns} of information retrieval and note writing compared to other expository writing setups~\citep{shen2023beyond}. Fine-grained information is needed to analyze and identify such patterns. 

We construct the \Dataset (\dataset) Dataset that aims to tease apart the ``reading''---retrieving relevant notes about the patient---and the ``writing''---synthesizing the findings into a coherent story---during the ED documentation process.
We work with a large urban academic medical center that has an audit logging system deployed to periodically save writing and reading activities.  
Thus far, in \dataset, we have collected granular reading and writing logs for a total of 48,192 written note versions and 6,514 visits for 5,803 patients over a period of 2 months.

Figure~\ref{fig:writing-timeline} shows an example of the ED documentation process. 

We break down the writing process into multiple \textbf{sessions}, which are intervals of time between two consecutive snapshots of the written note. 
We refer to each note opening as a \textbf{retrieval} action and link it with the sessions based on the time overlapping.\footnote{The opening of a note does not necessarily mean it is relevant and should be \emph{retrieved}---it can be discarded after a quick read by the doctor. However, since the total number of opened notes account for a small portion of the total available notes, the opening of a note provides important signal and we use it as an approximation for retrieval in our case.}
An ideal session should encapsulate both the retrieval of past notes and updates to the currently written note, (e.g., session 1 and 3 in the figure), with which we can understand how patient information is synthesized and incorporated in the written note. 
The periodic saving of the written note may also lead to sessions with only reading or writing (e.g., session 2 or 4), and the single action in these sessions can support better characterization of reading or writing.
During a patient's ED visit, there are 7.4 sessions on average. Each visit has 8 notes retrieved on average, and clinicians typically produce written notes that are 229 words long. Sessions that contain both reading and writing (14\% of the total sessions) have a median duration of 40 minutes. 

\subsection{Analyzing Reading and Writing Patterns in \dataset}

By comparing the read note text and written text within the same session, we can study how information is consumed and synthesized into the eventual written note.
For example, we can compute how much of the text is copied from the read document using ROUGE score~\citep{lin-2004-rouge}, a typical metric in document summarization.\footnote{In this setup, we can assume that the doctor's writing summarizes the read notes.} 
For each session, we can obtain the newly written text by comparing the current writing with text from the previous session. We calculate ROUGE between the new text and the concatenation of all read notes in the session. 
The average ROUGE-1 precision is 0.52 ($\pm0.26$), indicating that 52\% of the unigrams in the newly written text can be found in the read document. The ROUGE-2 and ROUGE-L precision are 0.12 ($\pm0.19$) and 0.40 ($\pm0.24$), respectively. 
The high ROUGE metrics indicate that there are frequent copies of the read document text and paraphrases, which aligns with the ``copy forwarding'' phenomena that is common during clinical documentation \citep{Tsou2017}.

\paragraph{Qualitative Insights on Reading Patterns}
We randomly select patients to gain insights on reading patterns over time specifically in the ED. 
For an example Patient 1 (Figure \ref{fig:clin_ex1}), the middle timeline shows the retrieval activity of multiple users during the ED visit. From the top left grid, we observe that multiple users will read the same note, which tends to be the most recent note available for the session where the user is reading the note. In Figure \ref{fig:clin_ex3} for Patient 3, the user who wrote the ED notes was an Internal Medicine resident, who read the same most recent note in three different sessions. Afterwards, the Internal Medicine attending read that same most recent note in two different sessions. We can hypothesize that the attending reviews the same past notes that the resident used to write the current note in order to validate the decision making. This pattern of reading the same notes over again suggests the potential benefit of incorporating a pinning functionality to revisit notes with high clinical value, which would save clinicians additional time searching for the same note again. 

In addition, we find that different users in the care team will have different information needs depending on their specialty and role. For instance, in Figure \ref{fig:clin_ex5} for Patient 5, the ED resident who wrote notes only read the most recent orthopedics note for a patient with left hip pain, which was all the context this user needed to proceed with patient care. In constrast, an orthopedics resident who read notes in a later session read not only the most recent orthopedics note, but also older orthopedics progress notes, as well as hospital medicine, nursing, and primary care notes to gather a more complete story about this patient. 
The differences in reading patterns reflect the differences in the users' roles: the ED resident must quickly understand the patient's most urgent needs to administer timely patient care, while an orthopedics resident acts as a consultant with more detailed expertise and bandwidth to comprehensively understand the patient's past medical history.

\section{Modeling Proactive Information Retrieval with \dataset}

\subsection{The Proactive Information Retrieval Task}

During the writing process, clinicians are facing high stress to capture all needed information in such a short time. An ED visit usually lasts 260 minutes, and doctors need to read 8 unique documents (each of length 626 words on average) multiple times all while seeing the patient and coordinating care with colleagues. Multiple clinicians on our team reported that they spend an excessive amount of time searching for potentially relevant notes in the EHR, and they still worry that key pieces of information might be neglected given the short period of time. They must also spend additional time reading the retrieved notes to gauge relevancy. Coupled with our findings, we choose to focus on the task of proactive information retrieval and leverage machine learning to demonstrate the utility of our framework.

\begin{figure}[hbt!]
    \centering
    \includegraphics[width=0.75\linewidth, ,height=0.75\textheight,keepaspectratio]{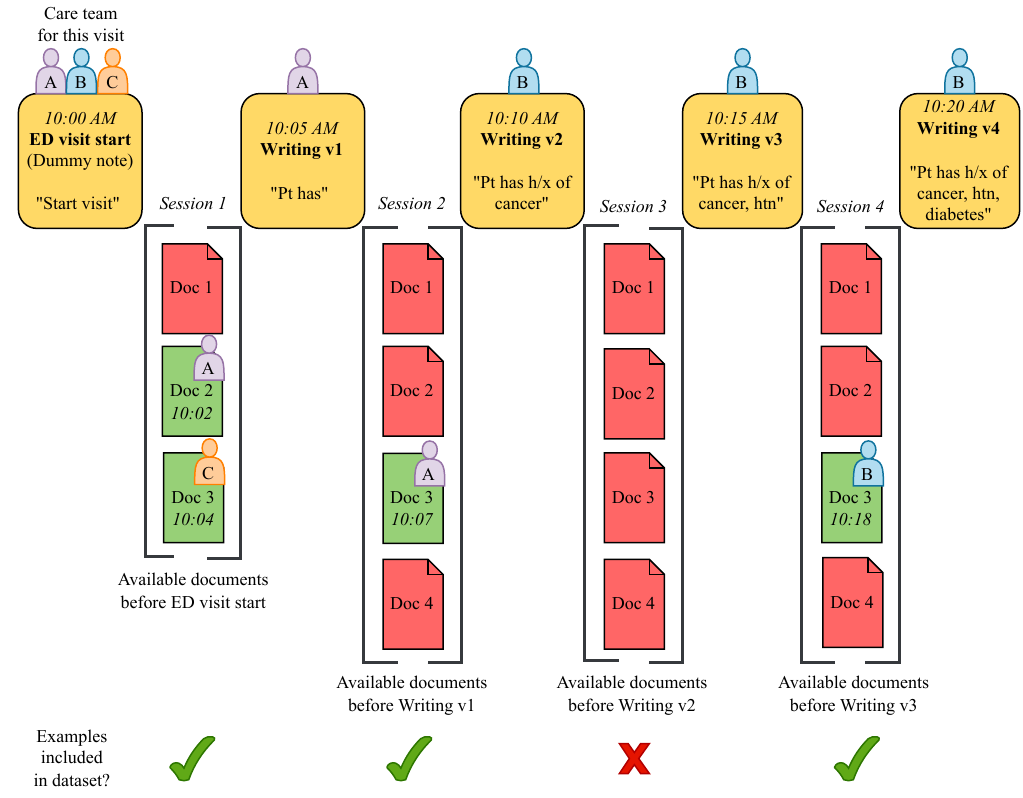}
    \caption{\small Dataset construction for the dynamic information retrieval task. Given a previously written note, this dynamic information retrieval task focuses on proactively predicting which available source documents are relevant to inform the next written note. The yellow blocks represent successive versions of written notes with timestamps and example text. The session for each written note is represented by the brackets preceding that note. For example, in the note writing session for Writing v1 (which we refer to as \textit{Session 1}), the available source documents are Documents 1, 2, and 3, which are created before the ED visit start time at 10:00 AM. In each note writing session, the green notes (positive label) are source documents that were read in the time since the previous written version and the next written version, and the red notes (negative label) were available but not read. In Session 1, Document 2 is labeled as positive since it is read at 10:02 AM, which is between the previous writing time at 10:00 AM and the next writing time at 10:05 AM. We define a dummy note at the ED visit start time to capture source documents read before users start writing note versions. The user next to a green or red note indicates the reader, and the user on top of the written note is the writer. Only notes from Sessions 1, 2, and 4 are included in the dataset because there is at least one positively labeled note in those sessions (the user is in an information seeking mode).}  
    \label{fig:task_negs_diagram}
    \vspace{-5mm}
\end{figure}

We consider proactive information retrieval as a \emph{binary classification task} of whether a source document\footnote{In the rest of the paper, we use the term source document and available document interchangeably to refer to documents that already exist in EHRs, as opposed to written ``notes'', which refer to the text that the clinicans are updating during the current visit.} should be retrieved in the next session given the current writing context. 
Formally, given the current written note $\mathbf{w}^{(i)}$, the model predicts $y_j^{(i+1)}\in\{0,1\}$ to denote whether to retrieve a source document $\mathbf{d}_j^{(i+1)}$ from a corpus of documents $\mathcal{D}^{(i+1)}$ available in session $s_{i+1}$.\footnote{We consider documents which are created before the start of session $s_{i+1}$ to be available for that session.}
Because sometimes new source documents will be produced in the middle of the writing, we constantly update the candidate document set $\mathcal{D}^{(i)}$ across sessions.
The model uses the text from the current written note $\mathbf{w}^{(i)}$ and source documents $\mathbf{d}_j$, as well as features from the patient visit and source document metadata. Figure \ref{fig:task_negs_diagram} describes the dataset construction for this task.

\subsection{Modeling Choices and Feature Construction}

In this work, we experiment with logistic regression models to predict whether to retrieve the source documents. 
Compared to other more complex machine learning methods, linear models' predictions are relatively easy to interpret and can provide valuable insights for understanding the data and the writing process. 
The models use a combination of the three types of features mentioned above; we summarize the features in Table~\ref{table:features-all} in the appendix, and we detail the preprocessing of these variables as follows. 

\paragraph{Patient and Clinician Information}

When a patient presents to the ED, a triage nurse records the chief complaint (a structured one to two word phrase that describes the patient’s medical status). 
We use chief complaint as the primary signal for the patient's context since is a critical component of ED decision making~\citep{Greenbaum2019}. 
It is used as categorical variable: there are a total of 421 different chief complaints across all ED visits in \dataset, and we choose to include the 100 most-frequent chief complaints, which covers 81.5\% of visits. 
We encode this feature into a one-hot vector, and we account for the fact that a patient can have multiple chief complaints.
We also include the role of the clinicians who are writing the notes to indicate their experience levels.

\paragraph{Source Document Metadata} 
Besides the textual information in the source document,  the source document metadata (e.g., the creation time, document type) can provide important signals for the models. 
For example, we observe that clinicians will prioritize reading documents of potential relevance that are created most recently. 
We thus include features based on documents' \emph{relative creation time} to the start of the current session and \emph{document recency}---rankings of available documents based on their creation time. 
Additionally, because the frequent retrieval of a document reflects strong signals of its relevance, we count how many times one document is read in previous sessions and add it to the feature set.
Other metadata such as service (e.g., oncology, cardiology) and note type (e.g., progress note, letter) are also included. For all the aforementioned features that are not categorical by nature, we group the data into intervals and convert them to one-hot vectors. 
Appendix~\ref{target_note_features_detail} details the distribution of the data and how we group features into categorical values.

\paragraph{Written Notes and Source Document Text}
We represent the text in both the source documents and written notes using a bag-of-words (BoW)-based method.  
The document is encoded as a vector that captures the presence of certain n-grams in the text. 
We include the top 20,000 unigrams that occur at least 10 times across the 
corpus, and add 500 bigrams and 500 trigram selected based on their Pointwise Mutual Information (PMI) score~\citep{Pecina2006, church1990word}.

We convert the original \dataset dataset into a training corpus of $\{(\mathbf{x}, y)\},$ where $\mathbf{x}$ is a concatenation of the feature vector described above, and the dimension of $\mathbf{x}$ is 38,989 when using all features.  
For each session in a patient's visit, we create $|\mathcal{D}^{(i)}|$ samples as the model generates a prediction $y_j^{(i)}$ per available source document $\mathbf{d}_j^{(i)}$.  
There are a total of 1.6 million total examples with 24,768 positive labels 
across 2,760 patients.
We partition the data into training (80\%) and testing (20\%) sets based on the patient IDs to avoid potential data leakage and use five-fold cross validation to tune regularization hyperparameters.

\subsection{Evaluation Approach}

For evaluation, we use traditional classification performance metrics in machine learning, including precision, recall, F1, and AUC. For precision, recall, and F1, we only report results for the positive label. We pay particular attention to the AUC, which is especially important given the class imbalance in the dataset. We interpret a higher AUC to mean that the classifier can better distinguish between relevant and non-relevant source documents across all classification thresholds. 

In addition, we consider a set of information retrieval metrics to assess how well the machine learning models surface source documents that satisfy the information needs for the next note writing session. The metrics of interest include \textit{precision@k}, \textit{recall@k}, and \textit{F1@k}. Precision@k indicates how many items in the top $k$ results were relevant, recall@k indicates how many actual relevant results were shown in the top $k$ out of all actual relevant results for the query, and F1@k is the harmonic mean of precision@k and recall@k. For all classification and retrieval metrics, the minimum is 0 and maximum is 1, where a higher score indicates a more accurate model. 

Based on clinician input, we choose $k=10$ to present a wide range of possibly relevant notes and help the users ensure they have reached information saturation (i.e., all relevant information has been collected and the user does not want to read any more notes). Clinicians often tend to build patients' stories in a chronological manner to understand the reasons behind recent symptoms and medical decisions, so the order in which they read surfaced notes is important for evaluation of our prediction. 
Thus, for evaluation, the notes with the top 10 predicted probabilities are displayed chronologically.
Source documents are ranked such that the most recently written note that is predicted to be relevant is ranked higher. We also include a baseline to rank the 10 most recent notes per session to ground our quantitative evaluation.

\subsection{Explicit Relevance Judgments from Domain Experts}

To augment our performance evaluation, we conducted a chart review of five randomly selected patients with two practicing physicians, one board-certified in Internal Medicine (``Doctor A") and one board-certified in Emergency Medicine (``Doctor B"). Doctors from two different specialties were chosen in order to elicit a broader range of insights. Each doctor was provided with the age, sex, and chief complaint of the patient, and asked to review notes in the EHR as if they were the primary clinician seeing the patient in the ED at the time. The two doctors were asked to ``think aloud" as they navigated through the EHR in order to explain why they chose to read some notes and ignore others. For each note they chose to read, they were asked to categorize it using one of three labels: \textit{relevant positive} (the note contained information directly helpful to the patient's current presentation), \textit{relevant negative} (the note contained information that is helpful to know but not directly helpful to the patient's current presentation), and \textit{irrelevant negative} (the note contained information that is not helpful to know and is not directly helpful to the patient's current presentation). All notes that were not retrieved were also labeled as irrelevant negative. This process continued for subsequent written note versions for each patient, where the doctor was asked which notes they would surface after receiving more information from the most recent written note version. We use these explicit labels (Table \ref{tab:explicit_judge}) to compare if the notes a clinician wants to read match the predicted notes from the models.

\section{Results}

In this section, we report quantitative performance metrics to evaluate our machine learning models. We examine the classification and retrieval performance, as well as conduct an interpretability analysis and ablation study. From a qualitative perspective, we choose a subset of six random patient visits to perform a deep dive on how well the predicted notes reflect actual clinical relevance. We augment our qualitative analysis by evaluating the relevance of predicted notes using independently collected explicit judgments by two clinicians for five of the six patient visits from the preceding analysis. 

\subsection{Quantitative Results}

\begin{table}[t]
  \centering 
  \caption{Classification evaluation metrics for machine learning models on held out data.}
  \begin{tabular}{lllllll}
  \toprule
    \textbf{Task} &\textbf{Model} & \textbf{Precision} & \textbf{Recall} & \textbf{F1} & \textbf{AUC} \\
    \midrule

    Original & Chief complaint only & 0.0&0.0&0.0&0.581\\
    
    Original & Baseline & 0.637 & 0.279 & 0.388 &0.852\\
    
    Original & Metadata only& 0.694 & 0.361 & 0.475 &0.962\\

    Original & Metadata+BoW & 0.719 & 0.369 & 0.488 &0.963\\
    
    \midrule
    
    Ablation & Metadata only& 0.418 & 0.027 & 0.050 &0.906\\
    Ablation & Metadata+BoW & 0.432 & 0.037 & 0.069 &0.916\\
    
    \bottomrule
  \end{tabular}
  \label{tab:aucs_orig} 
\end{table}

\begin{table}[t]
    \centering 
    \caption{Information retrieval evaluation metrics of predicted relevant notes.}
    \begin{tabular}{llll}\toprule
    
                \textbf{Model} & \textbf{Precision@10}  & \textbf{Recall@10} & \textbf{F1@10} \\\midrule

    Baseline& 0.496 & 0.342 & 0.706\\
    
    Metadata only & 0.698 & 0.391 & 0.457\\
    
    Metadata+BoW & 0.720 & 0.394 & 0.455\\

    \bottomrule
    \end{tabular}
  \label{tab:ir_metrics} 
\end{table}

The traditional metrics evaluating classification performance of the machine learning models for the prediction task are shown in the upper rows of Table \ref{tab:aucs_orig}. The classification performance using only chief complaint is poor as expected, since this model does not encode any information about the source document. We observe marked improvements in predictive performance for the metadata only model and metadata+BoW model. 

Table \ref{tab:bow_features_orig} displays the top 20 most important positively and negatively weighted features for the metadata+BoW model. From these weightings, we observe that the most important features to predict a positive label are related to the source document's recency: a more recent note (e.g., the most recent note or among the most recent 3 notes) means that the source document is more likely to be read, while a less recent note (e.g., older than the 10 most recent notes) indicates the note will be less relevant. 
In addition, initial note types are predictive of relevant notes, while letters are predictive of less relevant notes. This finding aligns with clinical intuition, as initial notes tend to be written when the ED consults a different specialty or a patient starts to see a specialist for a specific condition. This table also shows that source documents that have not been read before are not likely to be relevant for the current writing session. 
The most predictive written and source words are more difficult to interpret since they are unigrams; however, we can infer that 
source documents containing the ``ed" term may provide more context to the patient's current ED visit. In contrast, a source document containing the ``nursing" term may not be as relevant, since nursing notes generally track daily accounts of patient vitals and status, as opposed to critical diagnoses and treatments.

Table \ref{tab:ir_metrics} shows the information retrieval metrics for the machine learning models on the original task. Out of all these metrics, the precision@10 is the most relevant in evaluating the clinical utility of our prediction models. For the best model, the precision@10 is 0.720. This indicates that a majority of the 10 relevant suggestions displayed to a clinician in the EHR interface would be relevant to read. 

\paragraph{Ablation} To better understand the features that drive the model's predictions, we perform an ablation on the source document recency. The source document's recency feature significantly impacts the prediction, so we remove all data points where the source document is within the most recent 5 notes available for a note writing session. This will gauge the model's ability to select the older notes that still apply to the current context. The bottom rows in Table \ref{tab:aucs_orig} show the classification performance. For this ablation, the AUC of the metadata+BoW model drops from 0.963 to 0.916, indicating that the recency of the source document is highly predictive of its relevance. This aligns with clinical practice, as the most current updates to a patient's health are most relevant to the patient's current care. The difference in performance between the metadata only and metadata+BoW model is greater in the ablation compared to that of the original task, which suggests that there may be non-linear interactions between metadata and the note text.

Table \ref{tab:bow_features_ablate1} shows the most important positively and negatively weighted features for this ablation. Compared to the most important features from the original task, we find that the terms ``acute" and ``presenting" in the written notes are indicative of a user being in an information seeking mode (i.e., searching for relevant notes related to the current writing context), since these features have higher positive weights. Other notable positively weighted features include the ``imaging" and ``exam" terms in source documents, which may refer to results from CT scans or MRIs and physical exam results, respectively. The results from imaging and physical exams often provide useful historical information for clinicians. Finally, the chief complaint ``Transfer" has a relatively high positive weight, which suggests that a vague chief complaint prompts users to search for more information to contextualize why a patient is being transferred to a different specialty.

\begin{table}[t]
    \centering 
    \caption{Most important positive and negative features for the metadata+BoW model for source documents older than the 5 most recent available. The suffix \textit{\_s} indicates the word was in the source document, and the suffix \textit{\_w} indicates the word was in the written note. Source document services are denoted by the suffix \textit{\_serv}, and note types are denoted by the suffix \textit{\_nt}.}
    \footnotesize
    \begin{tabular}{llllllll}\toprule
    \multicolumn{4}{c}{Positive weights} & \multicolumn{4}{c}{Negative weights}
    \\\cmidrule(lr){1-4}\cmidrule(lr){5-8}
    Feature   & Value&  Feature   & Value  & Feature & Value&  Feature   & Value\\\midrule

md\_s &	0.49 & 	well\_s &	0.13&   time\_diff\_more5yr 	&-2.08& 	Letter\_{nt} &	-0.20\\
time\_diff\_1mo &	0.36 & 	Transfer &	0.13& 	recency\_{more10} &	-1.70& 	Progress note\_{nt} &	-0.18\\
Initial note\_{nt} &	0.32& 	acute\_w &	0.13& 	read\_before\_0 &	-1.67& 	code\_s &	-0.17\\
language\_s &	0.28& 	tone\_s &	0.13 & 	time\_diff\_5yr &	-1.35& 	rehab\_w &	-0.17\\
incontinence\_w &	0.25& 	resident\_s &	0.13& 	affiliation\_s &	-0.58& 	interventions\_s &	-0.15\\
Neurology\_{serv} &	0.22& 	family\_w &	0.12& 	at\_s &	-0.45& 	multiple\_w &	-0.15\\
Psychiatry\_{serv} &	0.21& 	ed\_s 	&0.12& 	signed\_s &	-0.32& 	esrd\_w &	-0.14\\
imaging\_s &	0.14& 	hpi\_s &	0.12& 	tm\_s 	&-0.27& 	hpi\_w &	-0.13\\
exam\_s &	0.14& 	[DATE]\_s &	0.12 & 	nursing\_s &	-0.25& 	subjective\_s &	-0.13\\
dr\_s	&0.13& 	presenting\_w &	0.11& 	since\_w 	&-0.20& 	dose\_s &	-0.12\\

    \bottomrule
    \end{tabular}
  \label{tab:bow_features_ablate1} 
\end{table}

\subsection{Qualitative Results}

For example Patient 1 (Figure \ref{fig:clin_ex1}), all read notes are captured within the top 6 predicted relevant notes for all sessions during the patient's visit. For 6 of the 8 sessions in this visit, the top ranked note was read, indicating that the predictive model surfaces notes with high clinical value to multiple users. The prediction model performs well in surfacing most recent notes, especially notes that are newly created since the last writing session. In the last session, the predicted rankings are updated to rank the most recently created notes at the top, which have direct relevance to the patient's care since they reflect clinical updates on the same day of the visit. The relevant positive notes that two clinician annotators wanted to surface during the first session were captured in the top 2 predicted rankings (Table \ref{tab:explicit_judge}). In this table, during the first session, the Internal Medicine physician (Dr. A) initially only surfaced the OBGYN telephone note as a relevant positive and did not surface any other notes. However, in a later session, she read a successive written note version and realized she missed a critical detail about the patient's past surgery. This prompted her to surface the surgery post operative check as a relevant positive. In contrast, the ED physician (Dr. B) read this post operative note during the first session, since he noticed that the first written note mentioned that the recent surgery had occurred.

Next, consider Patient 2 (Figure \ref{fig:clin_ex2}) with a chief complaint of psychiatric evaluation. The two most recent psychology telephone notes were the only notes read and were ranked in the top 2, demonstrating ideal performance. Both clinician annotators agreed that these notes were relevant positive. While the two physician annotators agreed on the most high utility notes, the Internal Medicine physician annotator preferred to see more notes until she was confident that she collected all the relevant information needed. The Internal Medicine physician wanted to gain more context by reading older notes from the patient's primary care physician. Although she hypothesized these primary care notes might initially be relevant, they ultimately did not contain any useful information. The Internal Medicine physician annotator included an additional relevant positive note that was not read according to the audit logs, which was the most recent psychology progress note from 6 months prior to the visit start date. In contrast, the ED physician considered the psychology progress note to be a relevant negative, as it provided somewhat relevant context but ultimately did not change his decision making for the patient. Since he is more concerned with the patient's most recent mental state, a psychology progress note from 6 months ago will not affect his immediate actions. All notes read by both annotators (regardless of their label) are displayed in top 10 predicted ranking, thus facilitating both personalities. The Internal Medicine physician is able to see a wider range of possibly relevant notes to gain confidence in all potential medical hypotheses, while the ED physician is able to quickly retrieve the most relevant notes at the top of the ranked list and forgo reading the remaining suggestions.

For Patient 3 (Figure \ref{fig:clin_ex3}), all read notes are captured within the top 6 predicted relevant notes for all sessions during the patient's visit. According to the explicit judgments, the two notes read by Dr. A were captured in the top 2 predicted notes, while the four notes read by Dr. B were captured in the top 4 predicted notes. These results show not only strong predictive performance, but also support using a chronological ranking for a more user-friendly EHR interface. Both clinician annotators read the relevant positive notes in the same order as that of the predicted rankings. 

Patient 4 (Figure \ref{fig:clin_ex4}) had a complicated medical history involving relevant notes from a variety of services, as well as a vague chief complaint of ``Transfer". Although a majority of read notes were surfaced by the predicted rankings, several read notes were not. The notes not captured by the rankings were older notes from months or years ago, which contained information related to conditions with long lasting effects (e.g., a transplant performed a decade ago). The relatively poorer performance of the predictions for this one patient indicates the need for future work to focus on non-linear methods such as language modeling to better capture the semantic relationships between written notes and older source documents. However, compared with the explicit judgments, all desired relevant positive notes from Dr. A were surfaced in the top 10, and only one relevant positive note read by Dr. B was not surfaced in the predicted rankings. Dr. B explained that he wanted to surface the Infectious Diseases note (which was not captured in the rankings) because he knew that this service has a track record of providing excellently written, comprehensive notes. This insight from clinical experience was not contained in the chief complaint or other metadata, but should perhaps be incorporated as a feature in the next iteration of our predictive model.

In Patient 5's visit (Figure \ref{fig:clin_ex5}), we observe that the predictive model correctly ranks note \#66 in the top 10 relevant suggestions, which is not in the 10 most recent notes available for the visit. This could be because that note is in the Hospital Medicine service, as opposed to the more common services found in the 10 most recent available notes (e.g., Orthopedics and Primary Care). From a clinical perspective, the last note from when the patient is hospitalized can provide relevant information as to why the patient was discharged. In the last session, although there were two notes that were read and not ranked in the top 10 predictions, the top 2 predictions completely align with the labels from both clinician annotators. Both annotators took notice of Patient 5's chief complaint of left hip pain and immediately surfaced the most recent Orthopedics note from the same day as the visit. The chief complaint often serves as a useful signal to direct the user to look for notes in a particular specialty. This note was predicted first, followed by the relevant negative note that both annotators agreed on: an older orthopedics progress note from 1 month ago. Although the top 10 predicted rankings did not capture notes \#4 and \#27 that were read by the orthopedics resident, the ED resident and both annotators only deemed note \#1 to be the only relevant positive. This note's position at the top of the rankings achieves our model's intended performance. With a closer inspection of the notes missed in the rankings, we find that note \#4 (a nursing progress note) is read for only 23 seconds before reading note \#27 (an older orthopedics progress note). We can hypothesize that User B quickly read note \#4 and did not find any useful information, so User B continued the search for other possibly useful notes.

\begin{figure}[hbt!]
    \centering
    \includegraphics[width=\linewidth,height=\textheight,keepaspectratio]{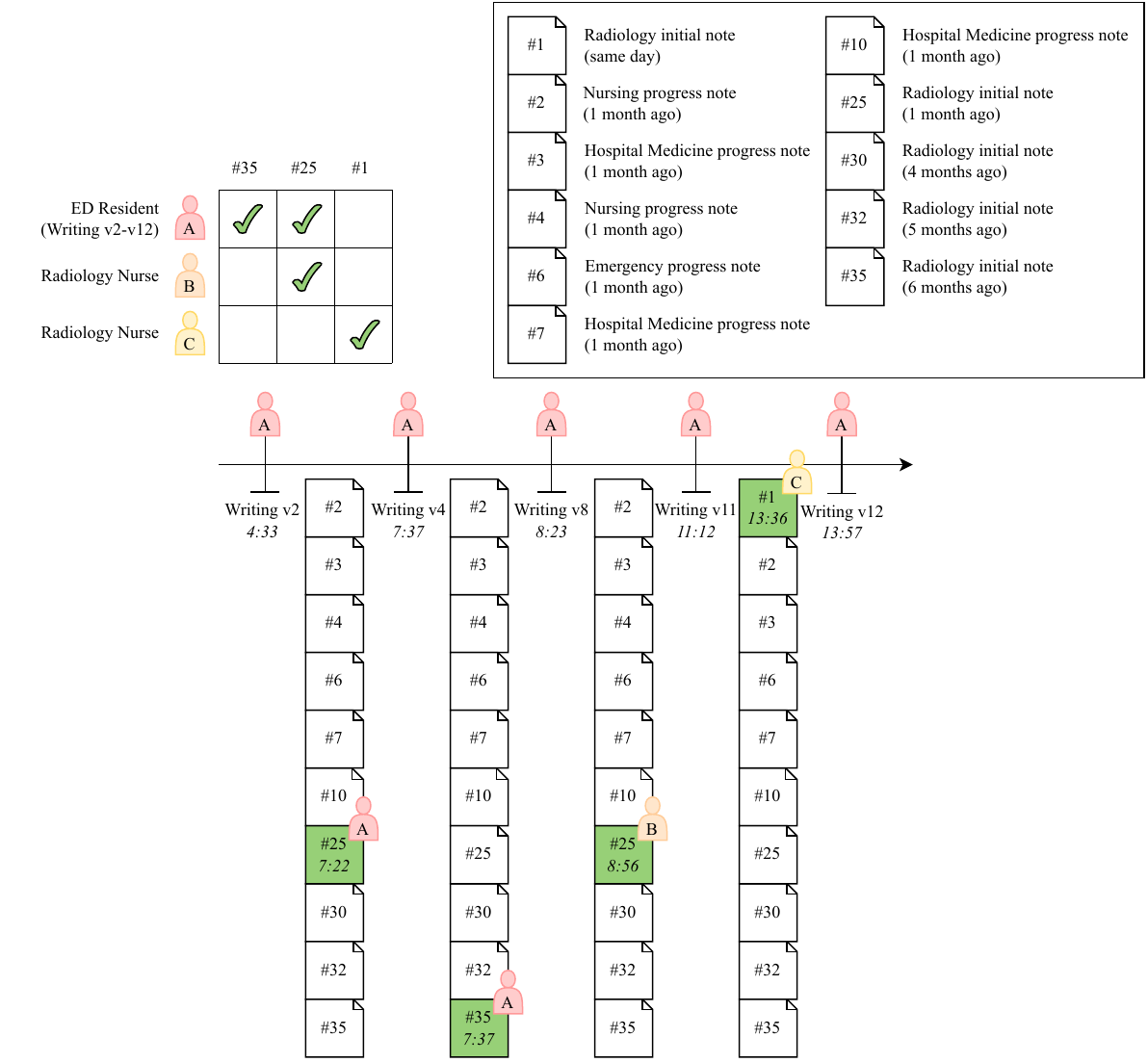}
    \caption{\small Deep dive qualitative analysis for example Patient 6 with a chief complaint of gastrostomy tube evaluation. The timeline (bottom) shows the top 10 predicted relevant notes in ranked order for each session. The green notes are predicted relevant and actually read. The italicized time is when the note was actually read. The grid (top left) indicates which user read which notes during the visit. The source documents are numbered such that the lower number indicates a more recently created note (e.g., note \#1 is the most recent note available for the visit). The top right box shows the ranked documents' service, note type, and time relative to the visit start.}
    \label{fig:clin_ex6}
    \vspace{-6mm}
\end{figure}

Patient 6's visit (Figure \ref{fig:clin_ex6}) also shows instances where the predictive model correctly ranks notes in the top 10 relevant suggestions, despite not being in the 10 most recent notes available for the visit. The patient's chief complaint of gastrostomy tube evaluation can explain why older notes are correctly predicted. A gastrostomy tube describes a tube inserted into the stomach under radiological guidance to facilitate nutrition and fluid movement. Despite being less recent, read notes \#25 and \#35 are correctly surfaced because they are created by the radiology service. Through an interpretability analysis, we find that the most important weights for these read notes are the terms ``MD", ``ED", ``Dr.", ``cosigned", ``addendum", ``consult", and ``evaluation" for note \#25 and \#35. In addition, the initial note type feature for both notes had a high predictive weight for the positive label. Upon closer inspection of the text, these notes were both radiology consult notes for a patient who was referred to the ED for evaluation of a gastrostomy tube. The notes were cosigned by an attending physician, who wrote an addendum confirming the decision making.

\paragraph{Overall Trends}

For all six visits, the predicted notes that were ranked higher in the list of 10 suggestions were indeed read by users. For all but one of the five visits that were evaluated with explicit judgments, the top ranked predicted note surfaced during the first session of the patient's visit was determined to be truly relevant (i.e., labeled as \textit{relevant positive}) by both clinician annotators. For three of these four visits, showing only the top ranked predicted note was sufficient information that both annotators needed to proceed with patient care. The annotators' relevant negative and irrelevant negative notes were also captured in the top 10 predicted notes for four out of five visits. Overall, the model performs quite well in predicting the notes the annotators and other users definitely want to read. The predicted notes ranked lower on the list of 10 suggestions also reflect the notes the clinicians might want to read.

The two clinician experts read different notes in the EHR to arrive at the same conclusion about each patient’s care. We hypothesize that different clinicians of the same specialty may navigate information in the EHR very differently, but are still seeking the same essential information that will help them effectively care for their patient. Our framework facilitates the process of finding this essential information, thus emphasizing its clinical utility.

\section{Discussion}

The amount of data in healthcare is growing exponentially, and clinicians need to efficiently retrieve and synthesize that information in order to provide the best care possible to their patients. However, much of this data exists in disparate formats and in different parts of the EHR, leading clinicians to spend more time digging for and documenting information rather than with their patients. We propose that machine learning can be used to create a dynamic information retrieval system that scours the chart on the clinician's behalf and suggests relevant data as the clinician learns more about the patient.

We present a novel framework for dynamic information retrieval of unstructured notes using EHR audit logs as a source of supervision for machine learning. We intentionally kept the machine learning setup straightforward to focus more on the design of the task. Our framing supports a natural followup using state of the art machine learning and NLP methods.  In this follow-up, we aim to understand the predictive
performance of pre-trained transformers such as BERT \citep{Devlin2019-rl} and ClinicalBERT \citep{alsentzer-etal-2019-publicly}
in note reading patterns as they relate to the writing process. These transformer
models have several self-attention layers \citep{Vaswani2017-zg} that will be able to capture complex relationships between the clinical context and the note text.

The clear win is that this system would allow clinicians to save time by finding relevant information more quickly. A well-designed system could also lead to better patient care by surfacing difficult-to-find information and suggesting patterns a clinician may not have initially deemed relevant, prompting the clinician to reassess a diagnosis they initially anchored on and evaluate potential biases.

\paragraph{Future Work}

We hypothesize that there may be three different phenotypes for information retrieval and documentation among clinicians that reflect (i) the inpatient setting, (ii) the outpatient setting, and (iii) the acute care setting. 
While we chose to focus our initial work on the emergency department setting (representing acute care) due to the need for rapid information retrieval with unfamiliar patients, we aim to expand to other clinical specialties in order to better understand how information retrieval needs vary across care settings. For example, a provider in oncology may see a medically complex patient every few weeks and want to surface information about more chronic issues. We also plan to further evaluate our predictive model through a pilot deployment, where we will ask users for real-time feedback during real patient encounters and investigate the effect of the deployment on patient turnaround times.

\acks{This work was supported by an award from the MIT Abdul Latif Jameel Clinic for Machine Learning in Health (J-Clinic), the National Science Foundation (NSF award no. IIS-2205320), the Machine Learning Core at Beth Israel Deaconess Medical Center, the MIT Deshpande Center, and the MachineLearningApplications@CSAIL initiative. The authors would like to thank Luke Murray for facilitating the research collaboration, as well as the reviewers and members of the MIT Clinical Machine Learning Group for their valuable feedback.}

\bibliography{reference}

\begin{thebibliography}{31}
\providecommand{\natexlab}[1]{#1}
\providecommand{\url}[1]{\texttt{#1}}
\expandafter\ifx\csname urlstyle\endcsname\relax
  \providecommand{\doi}[1]{doi: #1}\else
  \providecommand{\doi}{doi: \begingroup \urlstyle{rm}\Url}\fi

\bibitem[Adler-Milstein et~al.(2020)Adler-Milstein, Adelman, Tai-Seale, Patel,
  and Dymek]{Adler-Milstein2020-wz}
Julia Adler-Milstein, Jason~S Adelman, Ming Tai-Seale, Vimla~L Patel, and Chris
  Dymek.
\newblock {EHR} audit logs: A new goldmine for health services research?
\newblock \emph{J. Biomed. Inform.}, 101:\penalty0 103343, January 2020.

\bibitem[Agrawal et~al.(2022)Agrawal, Hegselmann, Lang, Kim, and
  Sontag]{agrawal2022large}
Monica Agrawal, Stefan Hegselmann, Hunter Lang, Yoon Kim, and David Sontag.
\newblock Large language models are few-shot clinical information extractors.
\newblock In \emph{Proceedings of the 2022 Conference on Empirical Methods in
  Natural Language Processing}, pages 1998--2022, 2022.

\bibitem[Alsentzer and Kim(2018)]{alsentzer2018extractive}
Emily Alsentzer and Anne Kim.
\newblock Extractive summarization of ehr discharge notes.
\newblock \emph{arXiv preprint arXiv:1810.12085}, 2018.

\bibitem[Alsentzer et~al.(2019)Alsentzer, Murphy, Boag, Weng, Jindi, Naumann,
  and McDermott]{alsentzer-etal-2019-publicly}
Emily Alsentzer, John Murphy, William Boag, Wei-Hung Weng, Di~Jindi, Tristan
  Naumann, and Matthew McDermott.
\newblock Publicly available clinical {BERT} embeddings.
\newblock In \emph{Proceedings of the 2nd Clinical Natural Language Processing
  Workshop}, pages 72--78, Minneapolis, Minnesota, USA, June 2019. Association
  for Computational Linguistics.
\newblock \doi{10.18653/v1/W19-1909}.
\newblock URL \url{https://aclanthology.org/W19-1909}.

\bibitem[Burton et~al.(2004)Burton, Anderson, and Kues]{burton2004using}
Lynda~C Burton, Gerard~F Anderson, and Irvin~W Kues.
\newblock Using electronic health records to help coordinate care.
\newblock \emph{The Milbank Quarterly}, 82\penalty0 (3):\penalty0 457--481,
  2004.

\bibitem[Church and Hanks(1990)]{church1990word}
Kenneth Church and Patrick Hanks.
\newblock Word association norms, mutual information, and lexicography.
\newblock \emph{Computational linguistics}, 16\penalty0 (1):\penalty0 22--29,
  1990.

\bibitem[Devlin et~al.(2019)Devlin, Chang, Lee, and Toutanova]{Devlin2019-rl}
Jacob Devlin, Ming-Wei Chang, Kenton Lee, and Kristina Toutanova.
\newblock {{BERT}}: Pre-training of deep bidirectional transformers for
  language understanding.
\newblock In \emph{Proceedings of the 2019 Conference of the North {A}merican
  Chapter of the Association for Computational Linguistics: Human Language
  Technologies, Volume 1 (Long and Short Papers)}, pages 4171--4186,
  Minneapolis, Minnesota, June 2019. Association for Computational Linguistics.

\bibitem[Gopinath et~al.(2020)Gopinath, Agrawal, Murray, Horng, Karger, and
  Sontag]{gopinath2020fast}
Divya Gopinath, Monica Agrawal, Luke Murray, Steven Horng, David Karger, and
  David Sontag.
\newblock Fast, structured clinical documentation via contextual autocomplete.
\newblock In \emph{Machine Learning for Healthcare Conference}, pages 842--870.
  PMLR, 2020.

\bibitem[Greenbaum et~al.(2019)Greenbaum, Jernite, Halpern, Calder, Nathanson,
  Sontag, and Horng]{Greenbaum2019}
Nathaniel~R Greenbaum, Yacine Jernite, Yoni Halpern, Shelley Calder, Larry~A
  Nathanson, David~A Sontag, and Steven Horng.
\newblock Improving documentation of presenting problems in the emergency
  department using a domain-specific ontology and machine learning-driven user
  interfaces.
\newblock \emph{International journal of medical informatics}, 132:\penalty0
  103981, 2019.

\bibitem[Hirsch et~al.(2015)Hirsch, Tanenbaum, Lipsky~Gorman, Liu, Schmitz,
  Hashorva, Ervits, Vawdrey, Sturm, and Elhadad]{hirsch2015harvest}
Jamie~S Hirsch, Jessica~S Tanenbaum, Sharon Lipsky~Gorman, Connie Liu, Eric
  Schmitz, Dritan Hashorva, Artem Ervits, David Vawdrey, Marc Sturm, and
  No{\'e}mie Elhadad.
\newblock Harvest, a longitudinal patient record summarizer.
\newblock \emph{Journal of the American Medical Informatics Association},
  22\penalty0 (2):\penalty0 263--274, 2015.

\bibitem[King et~al.(2019)King, Cooper, Clermont, Hochheiser, Hauskrecht,
  Sittig, and Visweswaran]{King2019-ja}
Andrew~J King, Gregory~F Cooper, Gilles Clermont, Harry Hochheiser, Milos
  Hauskrecht, Dean~F Sittig, and Shyam Visweswaran.
\newblock Using machine learning to selectively highlight patient information.
\newblock \emph{J. Biomed. Inform.}, 100:\penalty0 103327, December 2019.

\bibitem[Li et~al.(2022)Li, Pan, Goldwasser, Verma, Wong, Nuzumlalı, Rosand,
  Li, Zhang, Chang, Taylor, Krumholz, and Radev]{LI2022100511}
Irene Li, Jessica Pan, Jeremy Goldwasser, Neha Verma, Wai~Pan Wong,
  Muhammed~Yavuz Nuzumlalı, Benjamin Rosand, Yixin Li, Matthew Zhang, David
  Chang, R.~Andrew Taylor, Harlan~M. Krumholz, and Dragomir Radev.
\newblock Neural natural language processing for unstructured data in
  electronic health records: A review.
\newblock \emph{Computer Science Review}, 46:\penalty0 100511, 2022.
\newblock ISSN 1574-0137.
\newblock \doi{https://doi.org/10.1016/j.cosrev.2022.100511}.
\newblock URL
  \url{https://www.sciencedirect.com/science/article/pii/S1574013722000454}.

\bibitem[Li et~al.(2008)Li, Chase, Patel, Friedman, and Weng]{li2008comparing}
Li~Li, Herbert~S Chase, Chintan~O Patel, Carol Friedman, and Chunhua Weng.
\newblock Comparing icd9-encoded diagnoses and nlp-processed discharge
  summaries for clinical trials pre-screening: a case study.
\newblock In \emph{AMIA Annual Symposium Proceedings}, volume 2008, page 404.
  American Medical Informatics Association, 2008.

\bibitem[Lin(2004)]{lin-2004-rouge}
Chin-Yew Lin.
\newblock {ROUGE}: A package for automatic evaluation of summaries.
\newblock In \emph{Text Summarization Branches Out}, pages 74--81, Barcelona,
  Spain, July 2004. Association for Computational Linguistics.
\newblock URL \url{https://aclanthology.org/W04-1013}.

\bibitem[MacAvaney et~al.(2019)MacAvaney, Sotudeh, Cohan, Goharian, Talati, and
  Filice]{macavaney2019ontology}
Sean MacAvaney, Sajad Sotudeh, Arman Cohan, Nazli Goharian, Ish Talati, and
  Ross~W Filice.
\newblock Ontology-aware clinical abstractive summarization.
\newblock In \emph{Proceedings of the 42nd International ACM SIGIR Conference
  on Research and Development in Information Retrieval}, pages 1013--1016,
  2019.

\bibitem[Mamykina et~al.(2012)Mamykina, Vawdrey, Stetson, Zheng, and
  Hripcsak]{Mamykina2012-nb}
Lena Mamykina, David~K Vawdrey, Peter~D Stetson, Kai Zheng, and George
  Hripcsak.
\newblock Clinical documentation: composition or synthesis?
\newblock \emph{J. Am. Med. Inform. Assoc.}, 19\penalty0 (6):\penalty0
  1025--1031, November 2012.

\bibitem[Menachemi and Collum(2011)]{menachemi2011benefits}
Nir Menachemi and Taleah~H Collum.
\newblock Benefits and drawbacks of electronic health record systems.
\newblock \emph{Risk management and healthcare policy}, pages 47--55, 2011.

\bibitem[Moy et~al.(2021)Moy, Schwartz, Chen, Sadri, Lucas, Cato, and
  Rossetti]{moy2021measurement}
Amanda~J Moy, Jessica~M Schwartz, RuiJun Chen, Shirin Sadri, Eugene Lucas,
  Kenrick~D Cato, and Sarah~Collins Rossetti.
\newblock Measurement of clinical documentation burden among physicians and
  nurses using electronic health records: a scoping review.
\newblock \emph{Journal of the American Medical Informatics Association},
  28\penalty0 (5):\penalty0 998--1008, 2021.

\bibitem[Muhiyaddin et~al.(2022)Muhiyaddin, Elfadl, Mohamed, Shah, Alam,
  Abd-Alrazaq, and Househ]{Muhiyaddin2022}
Raghad Muhiyaddin, Asma Elfadl, Ebtehag Mohamed, Zubair Shah, Tanvir Alam, Alaa
  Abd-Alrazaq, and Mowafa Househ.
\newblock Electronic health records and physician burnout: a scoping review.
\newblock \emph{Informatics and Technology in Clinical Care and Public Health},
  pages 481--484, 2022.

\bibitem[Murray et~al.(2021)Murray, Gopinath, Agrawal, Horng, Sontag, and
  Karger]{murray2021medknowts}
Luke Murray, Divya Gopinath, Monica Agrawal, Steven Horng, David Sontag, and
  David~R Karger.
\newblock Medknowts: unified documentation and information retrieval for
  electronic health records.
\newblock In \emph{The 34th Annual ACM Symposium on User Interface Software and
  Technology}, pages 1169--1183, 2021.

\bibitem[Nygren and Henriksson(1992)]{nygren1992reading}
Else Nygren and Peter Henriksson.
\newblock Reading the medical record. i. analysis of physician's ways of
  reading the medical record.
\newblock \emph{Computer methods and programs in biomedicine}, 39\penalty0
  (1-2):\penalty0 1--12, 1992.

\bibitem[Pecina and Schlesinger(2006)]{Pecina2006}
Pavel Pecina and Pavel Schlesinger.
\newblock Combining association measures for collocation extraction.
\newblock In \emph{Proceedings of the COLING/ACL 2006 Main Conference Poster
  Sessions}, volume 2006, page 651–658. Association for Computational
  Linguistics, 2006.

\bibitem[Rule et~al.(2020)Rule, Chiang, and Hribar]{Rule2020-tu}
Adam Rule, Michael~F Chiang, and Michelle~R Hribar.
\newblock Using electronic health record audit logs to study clinical activity:
  a systematic review of aims, measures, and methods.
\newblock \emph{J. Am. Med. Inform. Assoc.}, 27\penalty0 (3):\penalty0
  480--490, March 2020.

\bibitem[Shen et~al.(2023)Shen, August, Siangliulue, Lo, Bragg, Hammerbacher,
  Downey, Chang, and Sontag]{shen2023beyond}
Zejiang Shen, Tal August, Pao Siangliulue, Kyle Lo, Jonathan Bragg, Jeff
  Hammerbacher, Doug Downey, Joseph~Chee Chang, and David Sontag.
\newblock Beyond summarization: Designing ai support for real-world expository
  writing tasks.
\newblock \emph{arXiv preprint arXiv:2304.02623}, 2023.

\bibitem[Shivade et~al.(2014)Shivade, Raghavan, Fosler-Lussier, Embi, Elhadad,
  Johnson, and Lai]{shivade2014review}
Chaitanya Shivade, Preethi Raghavan, Eric Fosler-Lussier, Peter~J Embi, Noemie
  Elhadad, Stephen~B Johnson, and Albert~M Lai.
\newblock A review of approaches to identifying patient phenotype cohorts using
  electronic health records.
\newblock \emph{Journal of the American Medical Informatics Association},
  21\penalty0 (2):\penalty0 221--230, 2014.

\bibitem[Sultanum et~al.(2019)Sultanum, Singh, Brudno, and
  Chevalier]{sultanumDoccurateCurationBasedApproach2019}
Nicole Sultanum, Devin Singh, Michael Brudno, and Fanny Chevalier.
\newblock {\emph{Doccurate}} : A {{Curation}}-{{Based Approach}} for {{Clinical
  Text Visualization}}.
\newblock \emph{IEEE Transactions on Visualization and Computer Graphics},
  25\penalty0 (1):\penalty0 142--151, January 2019.
\newblock ISSN 1077-2626, 1941-0506, 2160-9306.
\newblock \doi{10.1109/TVCG.2018.2864905}.

\bibitem[Tsou et~al.(2017)Tsou, Lehmann, Michel, Solomon, Possanza, and
  Gandhi]{Tsou2017}
Amy~Y Tsou, Christoph~U Lehmann, Jeremy Michel, Ronni Solomon, Lorraine
  Possanza, and Tejal Gandhi.
\newblock Safe practices for copy and paste in the ehr.
\newblock \emph{Appl Clin Inform.}, 8:\penalty0 12--34, January 2017.

\bibitem[Vaswani et~al.(2017)Vaswani, Shazeer, Parmar, Uszkoreit, Jones, Gomez,
  Kaiser, and Polosukhin]{Vaswani2017-zg}
Ashish Vaswani, Noam Shazeer, Niki Parmar, Jakob Uszkoreit, Llion Jones,
  Aidan~N Gomez, {\L}~Ukasz Kaiser, and Illia Polosukhin.
\newblock Attention is all you need.
\newblock In I~Guyon, U~V Luxburg, S~Bengio, H~Wallach, R~Fergus,
  S~Vishwanathan, and R~Garnett, editors, \emph{Advances in Neural Information
  Processing Systems}, volume~30. Curran Associates, Inc., 2017.

\bibitem[Wang et~al.(2018)Wang, Wang, Rastegar-Mojarad, Moon, Shen, Afzal, Liu,
  Zeng, Mehrabi, Sohn, et~al.]{wang2018clinical}
Yanshan Wang, Liwei Wang, Majid Rastegar-Mojarad, Sungrim Moon, Feichen Shen,
  Naveed Afzal, Sijia Liu, Yuqun Zeng, Saeed Mehrabi, Sunghwan Sohn, et~al.
\newblock Clinical information extraction applications: a literature review.
\newblock \emph{Journal of biomedical informatics}, 77:\penalty0 34--49, 2018.

\bibitem[Wei and Eickhoff(2018)]{wei2018embedding}
Xing Wei and Carsten Eickhoff.
\newblock Embedding electronic health records for clinical information
  retrieval.
\newblock \emph{arXiv preprint arXiv:1811.05402}, 2018.

\bibitem[Zheng et~al.(2009)Zheng, Padman, Johnson, and Diamond]{Zheng2009-jy}
Kai Zheng, Rema Padman, Michael~P Johnson, and Herbert~S Diamond.
\newblock An interface-driven analysis of user interactions with an electronic
  health records system.
\newblock \emph{J. Am. Med. Inform. Assoc.}, 16\penalty0 (2):\penalty0
  228--237, March 2009.

\end{thebibliography}

\appendix

\section{Details on Source Document Features}
\label{target_note_features_detail}

To encode the relative creation time of the source document as a feature, we first calculate the written note version creation time minus the source document creation time.
For every example in the dataset, this time difference is positive, since the available source document is created before the written note is created. For the \textit{relative creation time} feature $T$, a category $t \in T$ is defined such that the source document is written $t$ before the written note is created, where $t$ $\in \{$24 hours, 1 week, 1 month, 1 year, 5 years, more than 5 years$\}$. These buckets are non-overlapping, such that a source document will only fall in one of these categories. For example, a source document may be created within 24 hours (exclusive) to 1 week (inclusive) of the written note version. Table \ref{tab:time_bucket_dist_orig} shows the distribution of the \textit{relative creation time} feature for positive and negative examples. The time distribution for positive examples skews towards more recent notes, while the skew is towards older notes for negative examples.

We divide the \textit{document recency} feature into categories, such that the source document is in the most recent $r$ notes available for a note writing session, where $r \in \{1,3,5,10,>10\}$. For instance, if $r = 1$, then the source document is the most recent one available. The categories are non-overlapping, so if $r = 3$, then the source document was either the 2nd most recent or 3rd most recent available. Table \ref{tab:most_recent_dist} shows the distribution of the \textit{document recency} feature for positive and negative examples. Positive examples (source documents that are read) are more likely to be more recent, while negatively labeled source documents are more likely to be less recent relative to the patient's medical history at a specific point in time.

The \textit{read repetition} feature is defined such that a source document is read $p$ times in previous note sessions, where $p \in \{0, 1, \{2,3\}, \{4,5\}, >5\}$. 4,497 out of 11,557 total unique source documents (38.91\%) are read across multiple sessions. For source documents that are read across multiple sessions, the average number of times they are read is 3.94 (refer to Table \ref{tab:pin_stats} for a more detailed breakdown of statistics). We include this feature to better understand the longer-term dependence of source documents on subsequent written notes. Table \ref{tab:read_before_dist} shows the distribution of this feature for positive and negative examples.

\begin{table}[t]
    \resizebox{1.\linewidth}{!}{
    \begin{threeparttable}
        \caption{{\small The list of all available features and their dimensions in the model.}}
        \label{table:features-all}
        \small
        \begin{tabular}{lr}
        \toprule
        Feature Description                                                    & Dimensions \\
        \midrule
        \multicolumn{2}{c}{\textbf{Patient and Clinician Information}}                      \\
        \midrule
        Chief complaint(s) of the patient                                      & 100       \\
        Level of training (e.g., student or resident) of the clinician writer  & 3         \\
        \midrule
        \multicolumn{2}{c}{\textbf{Source Document Metadata}}                              \\
        \midrule
        Current Session Time $t_i$ - Source Document Creation Time             & 6         \\
        How recent the target note is out of all available notes for a session & 5         \\
        How many times the document is read in previous sessions               & 5         \\
        Clinical service of the document (e.g., oncology, cardiology, etc.)     & 80        \\
        Document type (e.g., progress note, initial note, telephone, etc.)      & 15        \\
        \midrule
        \multicolumn{2}{c}{\textbf{Source Document and Written Note Text}}                \\
        \midrule
        Bag of words features for the document text                            & 21,000     \\
        Bag of words features for the written text                             & 17,775     \\
        \bottomrule
        \end{tabular}
    \end{threeparttable}
    }
\end{table}

\begin{table}[t]
    \centering 
    \caption{Distribution of the \textit{relative creation time} feature for positive and negative examples. For time differences with a start and end time, the start time is exclusive and the end time is inclusive. In the first row, ``within 24 hours" is inclusive, and in the last row, ``more than 5 years" is exclusive.}
    \begin{tabular}{lcccccl}\toprule
    & \multicolumn{2}{c}{Positive} & \multicolumn{2}{c}{Negative}
    \\\cmidrule(lr){2-3}\cmidrule(lr){4-5}
                & Count  & \%    & Count  & \%\\\midrule
    
    Source note is within 24 hours of written note&2,082 & 17.88 & 1,338 & 0.10\\
    Source note is within 24 hours to 1 week of written note &1,790 & 15.37 & 6,761 & 0.53\\
    Source note is within 1 week to 1 month of written note &2,545 & 21.85 & 38,109 & 2.98\\
    Source note is within 1 month to 1 year of written note &3,774 & 32.40 & 286,921 & 22.41\\
    Source note is within 1 year to 5 years of written note &1,114 & 9.56 & 589,896 & 46.08\\
    Source note is more than 5 years from written note &342 & 2.94 & 357,026 & 27.89\\

    \bottomrule
    \end{tabular}
  \label{tab:time_bucket_dist_orig} 
\end{table}

\begin{table}[t]
    \centering 
    \caption{Distribution of the \textit{document recency} feature for positive and negative examples. The categories are non-overlapping.}
    \begin{tabular}{lcccccl}\toprule
    & \multicolumn{2}{c}{Positive} & \multicolumn{2}{c}{Negative}
    \\\cmidrule(lr){2-3}\cmidrule(lr){4-5}
                & Count  & \%    & Count  & \%\\\midrule
 
    Source note is the most recent note &2,934 & 25.19 & 2,313 & 0.18\\
    Source note is in the most recent 3 notes &2,538 & 21.79 & 7,103 & 0.55\\
    Source note is in the most recent 5 notes&1,325 & 11.38 & 8,016 & 0.63\\
    Source note is in the most recent 10 notes&1,479 & 12.70 & 21,080 & 1.65\\
    Source note is older than the most recent 10 notes&3,371 & 28.94 & 1,241,539 & 96.99\\

    \bottomrule
    \end{tabular}
  \label{tab:most_recent_dist} 
\end{table}

\begin{table}[t]
    \centering 
    \caption{Distribution of the \textit{read repetition} feature for positive and negative examples.}
    \begin{tabular}{lcccccl}\toprule
    & \multicolumn{2}{c}{Positive} & \multicolumn{2}{c}{Negative}
    \\\cmidrule(lr){2-3}\cmidrule(lr){4-5}
                & Count  & \%    & Count  & \%\\\midrule

    Source note read 0 times previously&8,552 & 73.43 & 1,272,316 & 99.40\\
    Source note read 1 time previously&2,015 & 17.30 & 6,350 & 0.50\\
    Source note read 2 or 3 times previously&934 & 8.02 & 1,256 & 0.01\\
    Source note read 4 or 5 times previously&130 & 1.12 & 114 & 0.01\\
    Source note read more than 5 times previously&16 & 0.14 & 15 & 0.001\\

    \bottomrule
    \end{tabular}
  \label{tab:read_before_dist} 
\end{table}

\begin{table}[t]
  \centering 
  \caption{Statistics for how often a source document is read across multiple sessions. 4,497 out of 11,557 total unique source documents (38.91\%) are read across multiple sessions.}
  \begin{tabular}{llllllll}
  \toprule
    \textbf{Subset} & \textbf{Average} & \textbf{Std dev} & \textbf{Min} &\textbf{25\%} & \textbf{50\%} & \textbf{75\%} & \textbf{Max} \\
    \midrule

       All notes &  2.14 &	2.45 &	1 &	1.0 &	1.0 &	2.0 	&45\\
       Read across multiple sessions & 3.94 &	3.46 &	2 &	2.0 &	3.0 &	4.0 	&45\\

    \bottomrule
  \end{tabular}
  \label{tab:pin_stats} 
\end{table}

\clearpage
\section{Feature Importances for Original Task}
\label{important_feats_orig}
\begin{table}[!htbp]
    \centering 
    \caption{Most important positive and negative features for the metadata+BoW model, including all source documents. The suffix $\_s$ indicates the word was in the source document, and the suffix $\_w$ indicates the word was in the written note. Source document services are denoted by the suffix $\_serv$, and note types are denoted by the suffix $\_nt$.}
    \footnotesize
    \begin{tabular}{llllllll}\toprule
    \multicolumn{4}{c}{Positive weights} & \multicolumn{4}{c}{Negative weights}
    \\\cmidrule(lr){1-4}\cmidrule(lr){5-8}
    Feature   & Value&  Feature   & Value  & Feature & Value&  Feature   & Value\\\midrule
    
recency\_1 &	1.37 & 	tone\_s 	&0.19&   recency\_{more10} & 	-2.64& 	cirrhosis\_w 	& -0.18\\
md\_s 	&0.44 & 	lactate\_w& 	0.16& 	time\_diff\_more5yr&  	-2.17& 	nursing\_s & 	-0.17\\
recency\_3 &	0.40& 	mental\_w &	0.15& 	time\_diff\_5yr & 	-1.45& 	spoke\_w 	& -0.17\\
time\_diff\_24hr& 	0.40& 	ed\_s 	&0.15& 	read\_before\_0 & 	-0.95& 	at\_s	& -0.16\\
Psychiatry\_{serv}& 	0.36& 	[DATE]\_s &	0.14& 	recency\_{10} & 	-0.64& 	signed\_s&  	-0.16\\
Neurology\_{serv} 	&0.33& 	previously\_w& 	0.14& 	affiliation\_s & 	-0.49& 	since\_w & 	-0.15\\
Initial note\_{nt} &	0.30& 	resident\_s &	0.12& 	time\_diff\_1yr & 	-0.38& 	tm\_s 	& -0.14\\
family\_w 	&0.21& 	abuse\_s &	0.12& 	on\_s & 	-0.26& 	including\_w & 	-0.14\\
Neurosurgery\_{serv}& 	0.21& 	dr\_s 	&0.12& 	subjective\_s & 	-0.24& 	mechanical\_s & 	-0.13\\
tox\_w 	&0.21& 	cosigned\_s& 	0.12& 	yes\_s 	& -0.18& 	Letter\_{nt} & 	-0.13\\

    \bottomrule
    \end{tabular}
  \label{tab:bow_features_orig} 
\end{table}

\clearpage
\section{Explicit Relevance Judgments}
\label{explicit_rele_judge_table}

\begin{table}[!htbp]
  \centering 
  \caption{Explicit relevance judgments from two domain experts: Dr. A is a board-certified Internal Medicine physician and Dr. B is a board-certified Emergency Medicine physician. The note labels include: relevant positive (RP), relevant negative (RN), and irrelevant negative (IN). Patient is the patient number corresponding to the examples shown in Appendix \ref{qual_examples}. These show only the notes that either Dr. A or Dr. B opened, so other notes available that were not opened are automatically irrelevant negative. The rank indicates where in the list of 10 predicted suggestions the note was ranked. An X indicates that the note was not predicted within the top 10. The time of the note creation in parentheses is relative to the ED start visit time for that patient.}
  \small
  \begin{tabular}{lllll}
  \toprule
\textbf{Expert} & \textbf{Patient} &  \textbf{Note description} & \textbf{Label} & \textbf{Rank}\\
    \midrule
    
    Dr. A & 1 &OBGYN telephone (same day) & RP & 1\\
    Dr. A & 1 &Surgery post operative check (3 days ago) & RP & 2\\
    \midrule
    Dr. B & 1 &Surgery post operative check (3 days ago) & RP & 2\\
    \specialrule{.2em}{.1em}{.1em} 

    Dr. A & 2 &Psychology telephone (1 day ago) & RP & 1\\
    Dr. A & 2 &Psychology telephone (2 days ago) & RP & 2\\
    Dr. A & 2 &Primary Care letter (1 mo ago) & IN & 3\\
    Dr. A & 2 &Primary Care telephone (1 month ago) & IN & 4\\
    Dr. A & 2 &Psychology most recent progress note (6 months ago) & RP & 7\\
    \midrule
    Dr. B & 2 &Psychology telephone (1 day ago) & RP & 1\\
    Dr. B & 2 &Psychology telephone (2 days ago) & RP & 2\\
    Dr. B & 2 &Psychology most recent progress note (6 months ago) & RN & 7\\
    \specialrule{.2em}{.1em}{.1em} 

    Dr. A & 3 &Oncology telephone (same day) & RP & 1\\ 
    Dr. A & 3 &Palliative Care telephone (same day) & RP & 2\\
    \midrule
    Dr. B & 3 &Oncology telephone (same day) & RP & 1\\
    Dr. B & 3 & Palliative Care telephone (same day) & RP & 2\\
    Dr. B & 3 &Palliative Care telephone (2 days ago) & RP & 3\\
    Dr. B & 3 & Palliative Care telephone (3 days ago) & RP & 4\\
    \specialrule{.2em}{.1em}{.1em} 

    Dr. A & 4 &Ophthalmology most recent note (2 weeks ago) & RP & 3\\ 
    Dr. A & 4 & Nephrology most recent note (2 weeks ago) & RP & 5\\ 
    Dr. A & 4 &Cardiology most recent note (3 weeks ago) & RP & 7\\ 
    Dr. A & 4  &Oncology progress note (1 month ago) & RP & 9\\ 
    \midrule
    Dr. B & 4 & Infectious Diseases transplant note (1 month ago) & RP & X\\
    Dr. B & 4 &Oncology progress note (1 month ago) & RN & 9\\ 
    \specialrule{.2em}{.1em}{.1em} 

    Dr. A & 5 &Orthopedics initial note (same day)  & RP & 1\\ 
    Dr. A & 5 &Orthopedics progress note (1 month ago)  & RN & 2\\ 
    \midrule
    Dr. B & 5 &Orthopedics initial note (same day)   & RP & 1\\ 
    Dr. B & 5 &Orthopedics progress note (1 month ago)& RN & 2\\

    \specialrule{.2em}{.1em}{.1em}

  \end{tabular}
  \label{tab:explicit_judge} 
\end{table}

\section{Qualitative Patient Examples}
\label{qual_examples}

\begin{figure}[!htbp]
    \centering
    \includegraphics[width=\linewidth]{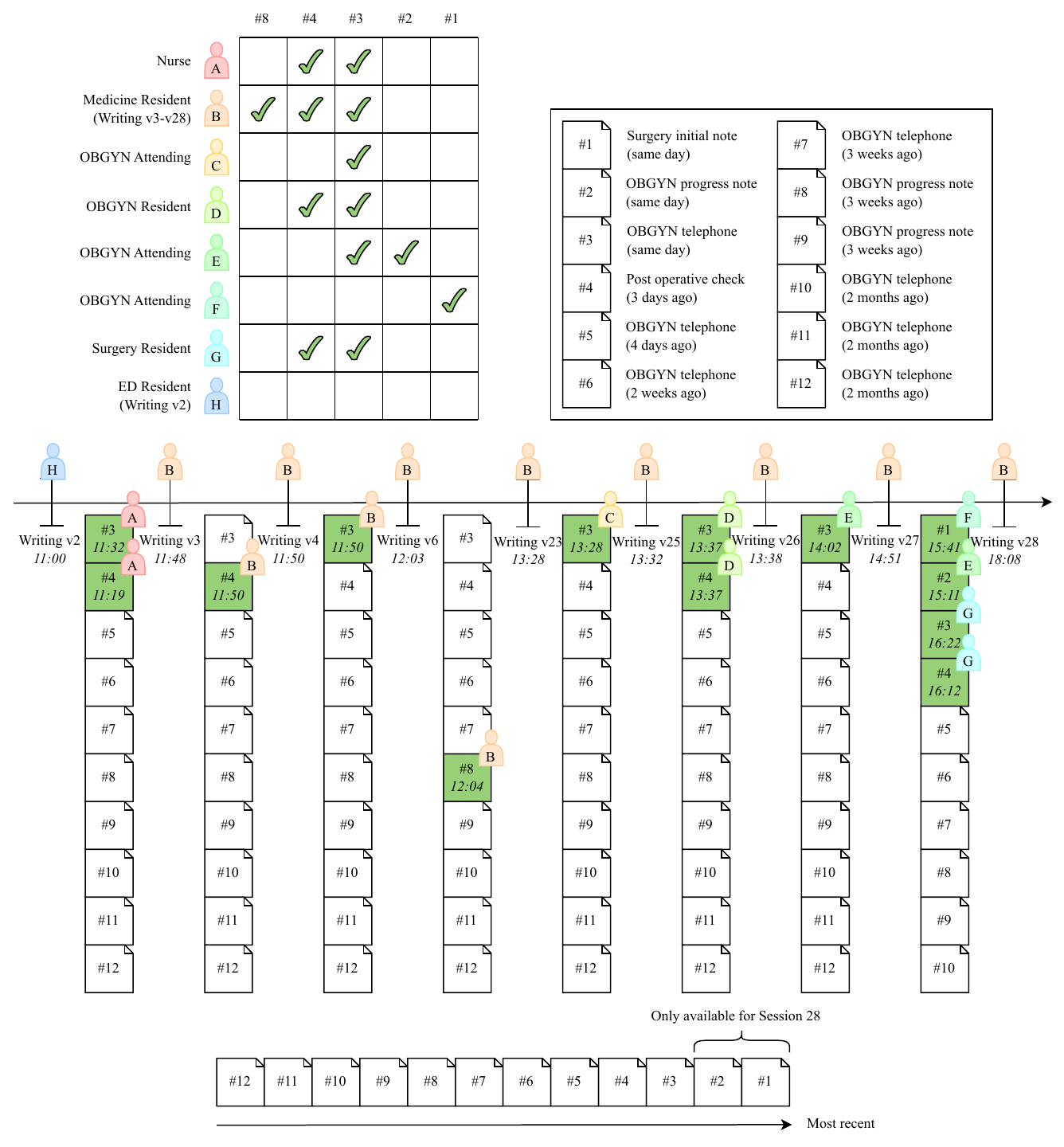}
    \small
    \caption{Deep dive qualitative analysis for example Patient 1 with a chief complaint of abdominal pain and fever. The timeline (middle) shows the top 10 predicted relevant notes in ranked order for each session. The green notes are predicted relevant and actually read. The italicized time is when the note was actually read. The grid (top left) indicates which user read which notes during the visit. The source documents are numbered such that the lower number indicates a more recently created note (e.g., note \#1 is the most recent note available for the visit). Observe that notes \#1 and \#2 were only available for Session 28 (the period before Writing v28 is written at time 18:08). The top right box shows the ranked source documents' service, note type, and time relative to the visit start.} 
    \label{fig:clin_ex1}
\end{figure}

\begin{figure}[t]
    \centering
    \includegraphics[width=0.5\linewidth,height=0.5\textheight,keepaspectratio]{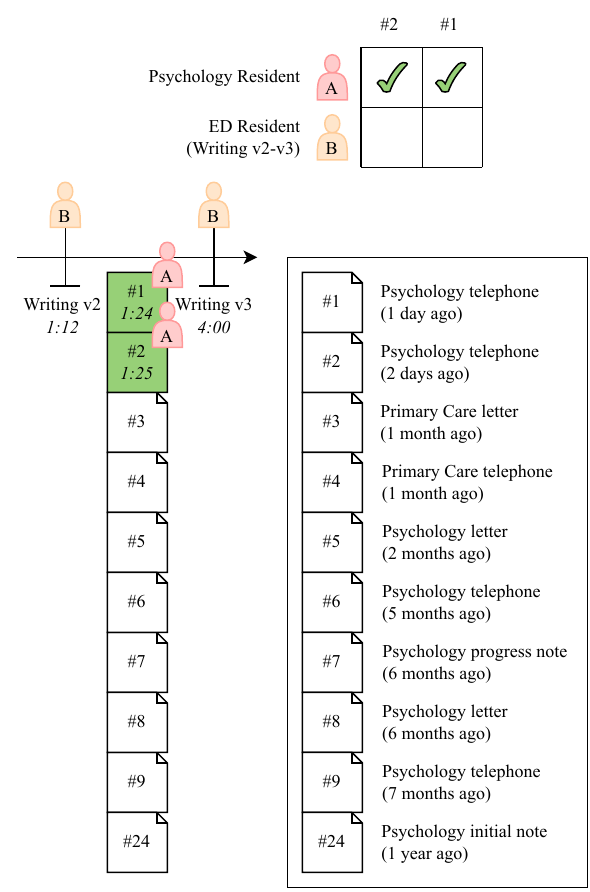}
    \caption{Deep dive qualitative analysis for example Patient 2 with a chief complaint of psychiatric evaluation. The timeline shows the top 10 predicted relevant notes in ranked order for each session. The green notes are predicted relevant and actually read. The italicized time is when the note was actually read. The grid (top) indicates which user read which notes during the visit. The source documents are numbered such that the lower number indicates a more recently created note (e.g., note \#1 is the most recent note available for the visit). The box on the right shows the ranked source documents' service, note type, and time relative to the visit start. }  
    \label{fig:clin_ex2}
\end{figure}

\begin{figure}[t]
    \centering
    \includegraphics[width=\linewidth]{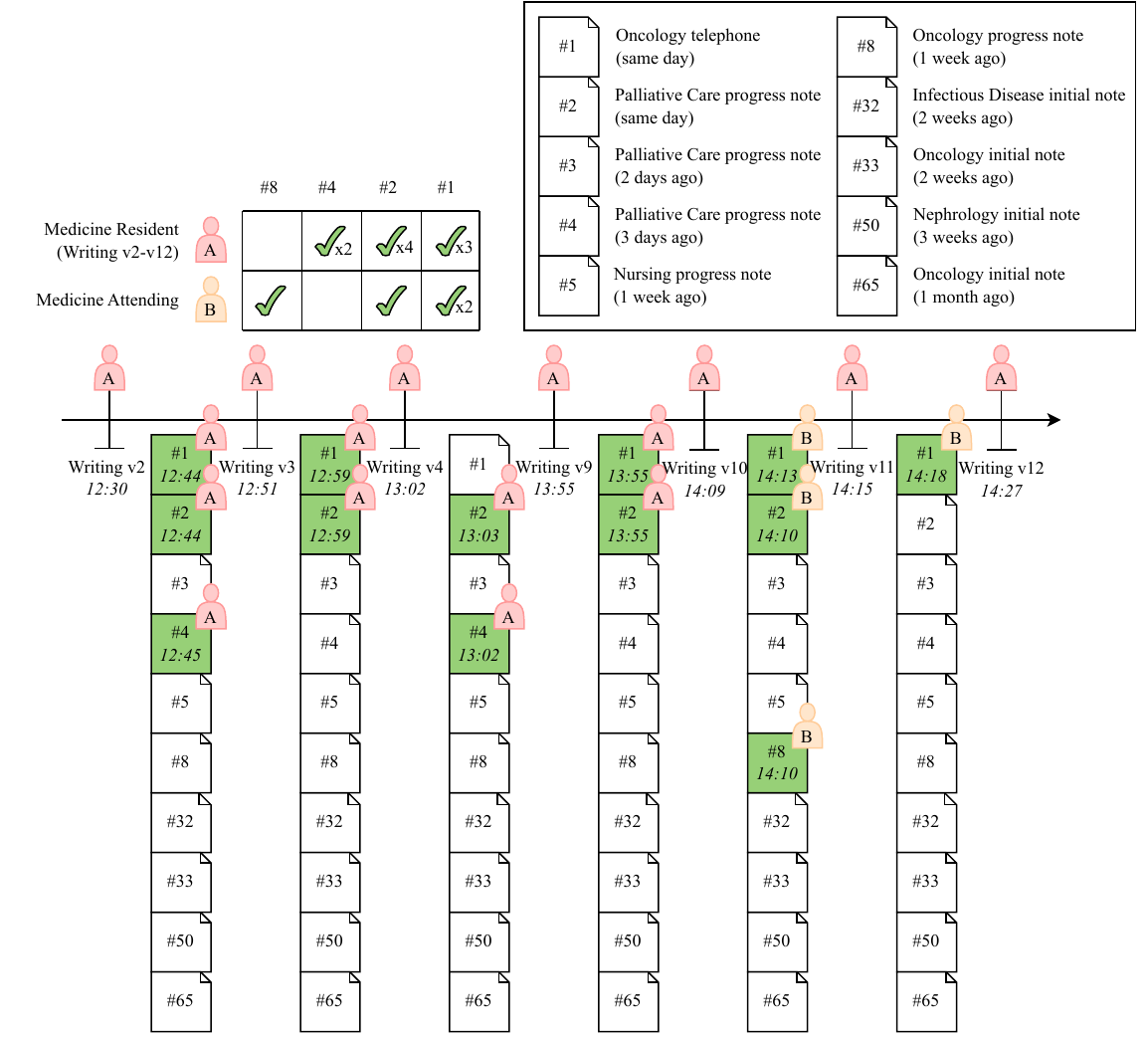}
    \caption{Deep dive qualitative analysis for example Patient 3 with a chief complaint of lethargy. The timeline (bottom) shows the top 10 predicted relevant notes in ranked order for each session. The green notes are predicted relevant and actually read. The italicized time is when the note was actually read. The grid (top left) indicates which user read which notes during the visit. In the grid, the multiplier next to the check mark for a user and note indicates that the note was read that many times in the visit. The source documents are numbered such that the lower number indicates a more recently created note (e.g., note \#1 is the most recent note available for the visit). The top right box shows the ranked source documents' service, note type, and time relative to the visit start.
    }  
    \label{fig:clin_ex3}
\end{figure}

\begin{figure}[t]
    \centering
    \includegraphics[width=\linewidth]{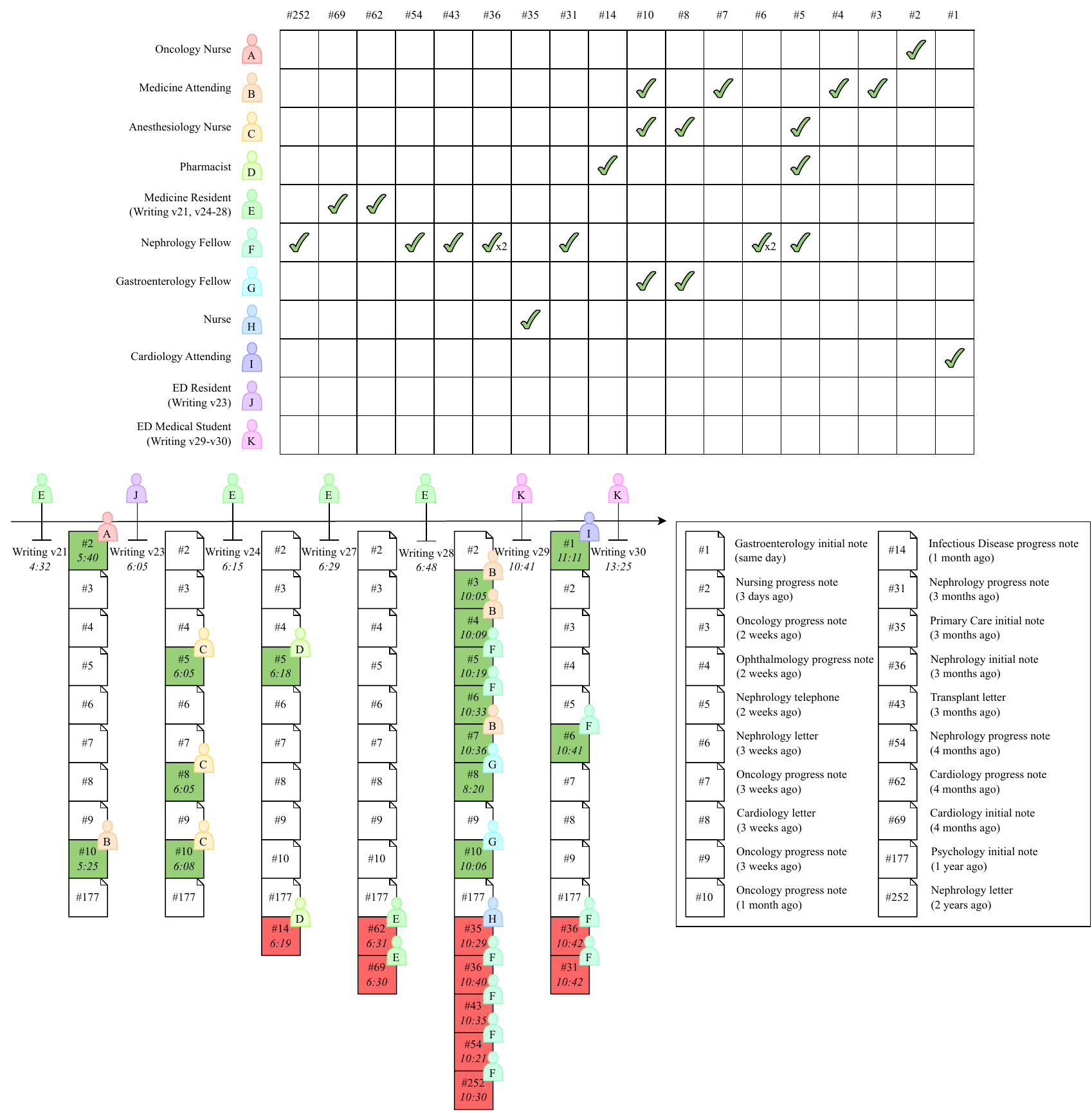}
    \caption{Deep dive qualitative analysis for example Patient 4 with a chief complaint of transfer. The timeline (bottom) shows the top 10 predicted relevant notes in ranked order for each session. The green notes are predicted relevant and actually read. The italicized time is when the note was actually read. Red colored notes are read during that session but not predicted in the top 10 rankings. The grid (top left) indicates which user read which notes during the visit. In the grid, the multiplier next to the check mark for a user and note indicates that the note was read that many times in the visit. The source documents are numbered such that the lower number indicates a more recently created note (e.g., note \#1 is the most recent note available for the visit). Source document \#1 was only available for Session 30. The top right box shows the ranked source documents' service, note type, and time relative to the visit start.}  
    \label{fig:clin_ex4}
\end{figure}

\begin{figure}[t]
    \centering
    \includegraphics[width=\linewidth,height=\textheight,keepaspectratio]{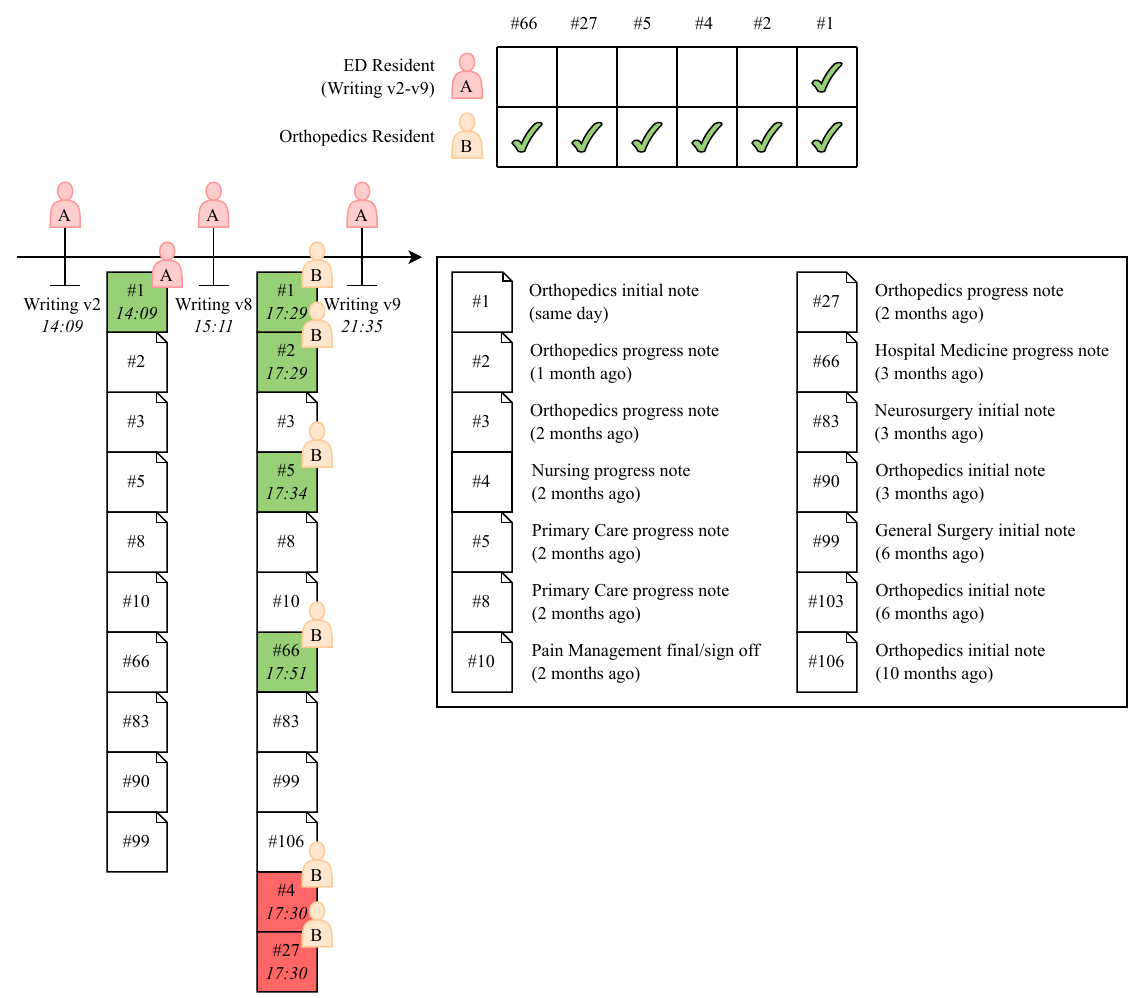}
    \caption{Deep dive qualitative analysis for example Patient 5 with a chief complaint of left hip pain. The timeline shows the top 10 predicted relevant notes in ranked order for each session. The green notes are predicted relevant and actually read. The italicized time is when the note was actually read. Red colored notes are read during that session but not predicted in the top 10 rankings. The grid (top) indicates which user read which notes during the visit. The source documents are numbered such that the lower number indicates a more recently created note (e.g., note \#1 is the most recent note available for the visit). The box on the right shows the ranked source documents' service, note type, and time relative to the visit start.}  
    \label{fig:clin_ex5}
\end{figure}

The predicted source document rankings and visit-specific context for Patient 1 are in Figure \ref{fig:clin_ex1}. The same information for Patients 2, 3, 4, and 5 are found in Figures \ref{fig:clin_ex2}, \ref{fig:clin_ex3}, \ref{fig:clin_ex4}, and \ref{fig:clin_ex5}, respectively.

\end{document}